\documentclass[10pt, conference]{IEEEtran}
\IEEEoverridecommandlockouts

\usepackage{color}
\usepackage{caption}
\usepackage{subcaption}

\usepackage{color}
\usepackage{enumerate}
\usepackage{booktabs}
\usepackage{ifthen}
\usepackage{graphicx}
\usepackage{balance}

\usepackage{url}
\newboolean{showcomments}
\setboolean{showcomments}{true}
\ifthenelse{\boolean{showcomments}}
{ \newcommand{\mynote}[2]{
    \fbox{\bfseries\sffamily\scriptsize#1}
    {\small$\blacktriangleright$\textsf{\emph{#2}}$\blacktriangleleft$}}}
{ \newcommand{\mynote}[2]{}}

\begin{document}

\pdfpagewidth 8.5in
\pdfpageheight 11.0in

\title{Harvesting Fix Hints in the History of Bugs}

\author{\IEEEauthorblockN{Tegawend\'e F. Bissyand\'e}
    \IEEEauthorblockA{SnT, University of Luxembourg, Luxembourg\\
	tegawende.bissyande@uni.lu\\
    }
}

\maketitle

\begin{abstract}
In software development, fixing bugs is an important task that is time consuming and cost-sensitive. While many approaches have been  proposed to automatically detect and patch software code, the strategies are limited to a set of identified bugs that were thoroughly studied to define their properties. They thus manage to cover a niche of faults such as infinite loops. We build on the assumption that bugs, and the associated user bug reports, are repetitive and propose a new approach of fix recommendations based on the history of bugs and their associated fixes. In our approach, once a bug is reported, it is automatically compared to all previously fixed bugs using information retrieval techniques and machine learning classification. Based on this comparison, we recommend top-{\em k} fix actions, identified from past fix examples, that may be suitable as hints for software developers to address the new bug.
\end{abstract}

\section{Introduction}\label{sec.intro}
Bug-free software is a myth~\cite{joyce89}. This is unfortunate since software is now pervasive in our lives, and history has established that software bugs can have extreme consequences including huge financial losses for businesses. Thus, fixing bugs is an important task in software development for continuously improving the dependability of software services. Fixing bugs is however time-consuming and cost-sensitive. Thus, constraints on resources, including time, manpower and testing environment, often lead project developers to release software that still contain many bugs~\cite{weiss07}. Consequently, software bugs are now everywhere and they have become critical both in terms of quantity (thousands new bugs per year in large projects [3]) and criticality (for safety or finance [4]). Necessarily, fixing bugs fast has become a primary concern for developers as well as users: e.g., the recent heartbleed\footnote{\url{http://heartbleed.org}} bug in the implementation of the popular OpenSSL library has left most systems vulnerable for about 2 years during which critical information was exposed.

Nonetheless, although the rapid pace of modern software development has produced numerous bugs that are reported back to developers for fixing, it should be acknowledged that development teams manage to fix and document an important number of these bugs. More importantly, as Andy Chou -- co-designer of the Coverity\footnote{\url{http://www.coverity.com} -- Coverity is currently the most successful commercial bug finding tool and was recently used to check for bugs in the Software of the Curiosity rover for its mission to Mars.} bug finding tool -- has stated about the deluge of buggy mobile device software, the exposed bugs are nothing new and are “actually well-known and well-understood in the development community - the same use after free and buffer overflow defects we have seen for decades”. Nevertheless, when project teams are flooded with bug reports, the time-to-fix interval increases and some bugs even remain unfixed for a very long time with all the risks that this entails. Already in 2006, a Mozilla developer has conceded that “every day, almost 300 bugs appear [...] far too much for only the Mozilla programmers to handle” [5, p. 363]. In such a context, without an automated tool for systematically analysing these bug reports and providing tips for how to fix them, most bugs will not be quickly dealt with, increasing the time-to-fix delays in the project. Some bugs, tagged as minor, will even go unnoticed.
To reduce the burden on software developers and limit the time-to-fix interval, a number of techniques have been proposed for automatically patching software buggy code ~\cite{Goues12, Perkins09,
Weimer09}. These approaches either focus on a limited set of well-known and easily identifiable faults such as infinite loops and obvious dereference faults for which they can infer a patch, or they infer patches by observing and comparing the execution of negative and positive test cases. However, such strategies cannot be applied to all projects. Furthermore, they cannot cover the large number of bug types that appear regularly in the life-cycle of a software program. Finally, current state-of-the art approaches on automated repair build on strong hypotheses, including the hypothesis that programs are accompanied with extensive test suites. The reality however is that adoption of software testing is limited~\cite{kochhar2013adoption,kochhar2013empirical}, and it is extremely challenging to have exhaustive test cases. Besides, software bugs are actually mostly identified by users and reported back to developers in hand-written bug reports. Indeed, software is too often shipped before testing can be completed.
When submitted to developers, bug reports can remain unattended for a very long time. Yet there are several reasons why one should fix bugs once they are reported. Indeed, it is well accepted3 that unfixed bugs can (1) camouflage other bugs, (2) suggest that quality is not important and (3) quickly become costly. E.g., Apple’s recent “goto fail” bug\footnote{\url{http://lwn.net/Articles/588369/}}, which was still unfixed 4 days after its disclosure, has been a major black eye for the company.

In this paper we introduce a new approach for drastically improving the fix rate of bugs through automatic bug fix recommandations. This approach leverages historical information extracted from software development artefacts including bug reports and bug fix links, to produce {\bf candidate fix actions} for newly reported bugs. We rely on 1) information retrieval techniques that have already been shown successful for helping bug triagers detect and dismiss duplicate bug reports, on 2) bug linking strategies to match past bug reports with their corresponding fixes, and on 2) semantic patch concepts to abstract, summarize and identify fix actions from concrete patches. We take these techniques further to accelerate, and if possible avoid, manual debugging by showing presenting developers with fix directions.

The contributions of this paper are as follows:
\begin{itemize}
\item we discuss fix recommendation and how it can yield positive results on software development processes.
\item we propose a new approach of fix recommendations using past experiences of bug fixes.
\item we discuss preliminary assessment on the feasibility of our approach.
\end{itemize}

The remainder of this paper is organized as follows: Section~\ref{sec:recommendation} discusses fix recommendations
and explore the different possibilities for a fix recommendation to be useful. Section~\ref{sec:approach} presents
our approach as well as samples of results from its implementation. Section~\ref{sec:related} discusses related work and 
Section~\ref{sec:conclusion} concludes.

\section{Related work}\label{sec:related}
Research on recommender systems for software repair (i.e., bug fixing) is part of a larger research agenda on software bugs. Currently, research is being done to find new bugs either statically (e.g. with abstract interpretation) or dynamically (e.g. with fuzz testing). There is also research to support the process of handling software bugs: how to automatically prioritize bug reports following criticality criteria? How to automatically assign bugs to competent developers? What are the code commits that are related/linked to a given bug report? Then, there is work on enhanced debugging which consist of approaches and tools to help localizing buggy pieces of code (fault localization), the causal dependencies (e.g. with program dependency graphs).

When it comes to actually repairing the bug, there have been contributions on recommending fixing actions~\cite{Jeffrey09} and even automatically repairing when the repair problem is stated in a certain way~\cite{Bartel2012} .  There is indeed today a momentum of automatic program repair, a research field where various approaches are devised to automatically fix programs once a fault is detected. Such approaches attempt to patch a program in a way that makes it pass all the tests in the test suite associated to the program. Weimer et al. use genetic programming for locating and repairing bugs in software written in C~\cite{Weimer09}. Their approach does not require formal specifications, program annotations or special coding practices. It uses standard test cases to detect bugs and assumes deterministic responses to test cases. Kim et al. have proposed to predefine templates of bug fixes using human-written patches~\cite{Kim13}. So far, and to the best of our knowledge, these approaches are not adopted in industry. Automated repair is indeed a young and immature research field, and current approaches still have a number of caveats:

(1) State-of-the-art techniques of automatic repair only perform on a limited set of fault types. Although empirical studies of big projects, such as of the Linux kernel development project~\cite{Palix:2011:FLT:1950365.1950401,Chou:2001:ESO:502034.502042}, show that a few common faults represent most of the faults, their categories vary in terms of complexity.

(2)  The proposed fixes are only temporary and may be out of tune with the rest of the code that is manually written by developers and that is thus in line with the project idiosyncratic programming rules.

(3) The performed fixes may only aim at avoiding immediate downtime of software usage. It does not however guarantee that the fix should be maintained, as it may not be the right fix in the spirit of developer intent, as pointed out by Monperrus~\cite{Monperrus2014}.

Other work, including approaches implemented within AutoFix-E~\cite{Wei2010, Pei:2011, Zeller14}, have proposed to leverage contracts inside programs to automatically patch program source code. Such approaches however often require a change in the coding habits within a development team. Furthermore, they still leave untouched many complex and/or unforeseen bugs which account for a large portion of bugs in some projects~\cite{Bissyande2012,bissyande2014ahead}.

Unfortunately, existing approaches have not managed to completely address the challenges of bug fixing in industry. Indeed, the industry standard remains to thoroughly review bug reports and manually write the corresponding fixes. Developers thus require new approaches and tools to help them readily understand bug report and infer the appropriate fix so as to reduce the time-to-fix delay.

\section{Automated Repair {\scriptsize vs} Fix Recommendation}
\label{sec:recommendation}
Software repair eventually comes down to delivering a patch that attempts to fix a deviated behaviour in software execution. In the context of automated repair, a primary goal of most patch generation systems is to enable applications to survive otherwise fatal inputs~\cite{qi2015analysis}. The produced patches in these cases are often quite simple where each repair consists typically of a condition check (e.g., NULL pointer validity) potentially followed by a simple action (e.g., halting function processing and returning). Because such fixes do not reflect the majority of bugs, researchers are now focusing on generating more realistic patches. Thus, some systems such as PAR~\cite{Kim13} rely on a set of human-learned patches to build fix templates. These approaches are still limited by the categories of templates considered.

In opposition to those approaches, we refer to as bug fix recommendation any hint addressed to a developer for fixing a reported bug. Bug fix recommendations therefore appear as an obvious workaround to problem of patch correctness associated with patch generation systems such as GenProg, RSRepair and AE. The main questions for recommender systems are thus 1) how the fix is inferred and 2) in which form it is presented to developers.

\begin{center}
	\begin{figure*}[!t]
	 	\centering
 	\includegraphics[width=0.9\textwidth]{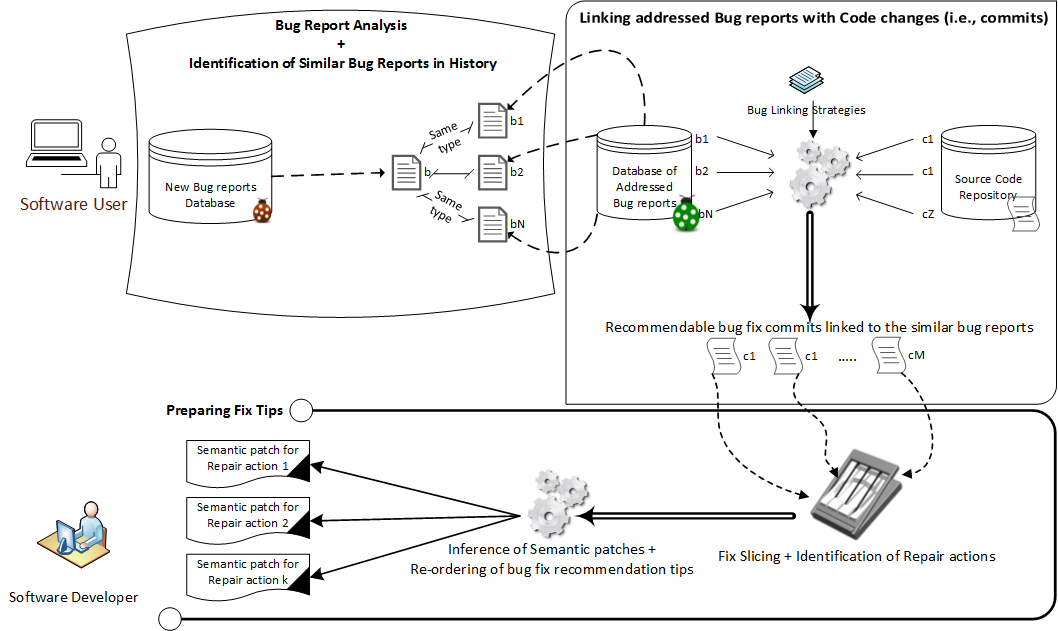}
 	\caption{Steps in the RECOMMEND approach}
 	\label{fig:recommend}
 \end{figure*}
\end{center}
 
\section{RECOMMEND}\label{sec:approach}
We propose a recommender system, named RECOMMEND, for bug fixing, to drastically improve the
fix rate of bugs by leveraging the history of bugs in (and potentially across) software development projects. Thus, with RECOMMEND we aim at enabling a better usage of the huge amount of user input that is submitted into bug trackers, in order to bring an improvement in (1) the fix rate and (2) the time- to-fix delay of user-reported bugs. Our approach will also contribute to make seamless the transitions between development teams of a software project, thus guaranteeing (3) the steady quality of patches over time. Indeed, by automatically finding similar bug reports and fixes from the past, we will help new developers understand faster how to fix a bug on a program whose programming idiosyncrasies are less obvious to them. In summary, RECOMMEND is designed as a robot helper to the human developer in his debugging activities. This robot drives a recommendation strategy contributing to:
\begin{enumerate}
	\item {\em faster repair of software}, preserving availability of software-related business services
	\item {\em coherent repair patches}, enhancing maintenance tasks, and improving productivity
\end{enumerate}

The RECOMMEND recommender system is devised through a mimicing approach of human scenario of bug fixing activities. In the next subsections we detail the different categories of these activities and present the corresponding step in RECOMMEND. Our approach unfolds in three steps during which the recommender system first attempts to capture the semantics of a bug report and associates it with previous bug reports, then infers the fix directions by simply obtained the fix links of the identified similar bug reports, before summarizing these fix directions into generic fix tips. Figure~\ref{fig:recommend} illustrates the different steps of the approach.

\subsection{Bug Report Understanding and Categorization}
In a software development project, when a bug report is filed, the first step of triaging is to understand what the bug is about so as to assign it to the right developer with the right skills corresponding to the problem~\cite{bissyande2013got}. This assignation can be performed following recorded tagged competences (e.g., the bug concerns sub-system A which is handled by developer x). However, previous work on bug management have proposed approaches to identified who should fix a given bug~\cite{Anvik06}. In RECOMMEND, this step can, to some extent, be automated building on the hypothesis that bugs, and thus bug reports, are repetitive. We thus leverage information retrieval techniques to cluster past bug reports in a project into categories to build an informal taxonomy of bug reports where each new bug report will be classified using supervised classifiers. For prototyping RECOMMEND we simply rely on existing topic modeling techniques, namely Latent Dirichlet Allocation (LDA)~\cite{lda} and Support Vector Machines (SVM)~\cite{svm}, to classify bug reporst based on topic similarity metrics. 

LDA is an unsupervised generative model that categorizes the words that appear in a corpus of documents into clusters, typically referred to as topics. By positing that words carry strong semantic information, LDA implements a statistical approach for discovering the latent semantic topics in collections of documents. The hypothesis here is that documents discussing similar topics will use similar groups of words. LDA has been used in the literature of bug report management, in particular to detect duplicate bug report. Generally, existing work also employ extra heuristics (such as date of report, submitter, bug status, etc.) to constrain the identification process to extremely similar (thus duplicate) reports.

SVM on the other hand is a supervised learning algorithm aimed at identifying patterns in data. A SVM model represents data samples as points in a multi-dimensional space. The training process ensures that samples of separate categories divided by a clear gap which is as wide as possible, leading to efficient binary classifiers.  

In this paper, we use the free-form textual description of the bug report as the document provided to the LDA computations. Before applying the LDA computation, we filter stop-words from the documents based on the Stopwords corpus provided by Porter et al. and included in the Natural Language Toolkit for Python programs. We also perform stemming, with the Porter algorithm~\cite{Porter:1997:ASS:275537.275705}, to reduce each word in its root, increasing the likelihood of matching words used by different bug reporter but in different inflections (e.g., {\em checked} and {\em checking} lead to the same stem {\tt check}). Figure~\ref{fig:topics_lda} lists the topics computed with the short descriptions of a sample of 100 randomly selected bug reports from the Linux kernel bugzilla repository\footnote{https://bugzilla.kernel.org}. The words in these topics correspond to the most recurrent inflected versions of all words sharing the same stem. These identified topics are promising: the first two topics reflect common kernel error messages related to the access to invalid memory.

\begin{figure}
{
\scriptsize
\begin{verbatim}
Topic 0: kernel unable handle request paging bug:
Topic 1: null pointer driver dereference connection check
Topic 2: oops problems processor interface modprobe bluetooth
Topic 3: kernel panic perf working _mm_pause blocking 
Topic 4: causes boot network ethernet vlan module 
Topic 5: fails memory toshiba_acpi high acer address 
Topic 6: issues eeepc failure related loopback problem
Topic 7: work packet using status bonding bridge 
Topic 8: enabled crashes running restart output system 
Topic 9: toshiba memory change reference available hangs
\end{verbatim}
}	
\caption{Topics Identified by applying LDA on the short description of 100 Linux bug reports}
\label{fig:topics_lda}
\end{figure}

Based on these topics to construct feature vectors and relying on a training set of bug reports from different categories directly related to empirically identified common faults in Linux~\cite{Palix:2011:FLT:1950365.1950401}, we can build SVM classifiers for categorizing new bug reports in those categories.

\subsection{Identification of Fix Examples}
\begin{center}
	\begin{figure}[!h]
	 	\centering
 	\includegraphics[width=1.3\linewidth]{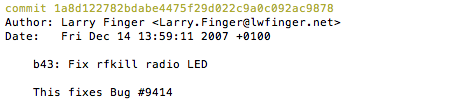}
 	\caption{A bug link explicited in the extract log of a code commit in Linux kernel repository}
 	\label{fig:bug_link}
 \end{figure}
\end{center}

In real world manual processes, once a bug is identified and localized, a developer must then craft the patch that will fix it. If he remembers having dealt with such a type of bug, he can immediately refer to the fix already applied and save time and effort. In RECOMMEND, we also attempt to automate this step. Since the previous step has already identified past bug reports that are ``of the same types'', their associated fixes can be leveraged as fix directions. The challenge is then to discover those associated fixes by matching the bug reports with code changes in the software project code base. This is known as bug linking in the literature~\cite{Bachmann:2010:MLB:1882291.1882308, Wu:2011:RRL:2025113.2025120}. For our prototype implementation we consider cases where such links are explicited directly in the code commit log as shown in the example of Figure~\ref{fig:bug_link}.

\subsection{Summarization of Fix Tips}
Directly presenting to developers top-k suitable fixes from the history of a project is not likely to help them write proper fixes for the bug at hand. Indeed, often commit patches often include fix actions as well as cosmetic changes or improvements unrelated to the bug reports. Furthermore, because so far in the description of RECOMMEND we have not considered including a bug localization approach to pinpoint the location to modify, providing complete examples of fixes may be confusing to the developer.

	Instead in our approach we consider slicing the fix examples found in previous step into fix actions and then to summarize them in higher-level generic patches. Slicing fixes into different actions statements will allow to rank fix actions by recurrence in the set of fix examples: we make the fair assumption that a recurring statement is indeed relevant to the bug at hand. High-level generic patches are patches that abstract over variable names from past fixes, and also, by construction, ignoring cosmetic changes. 
	
	To implement this approach we rely on the notion of semantic patches~\cite{padioleau08} specified in the Semantic Patch Language (SmPL) which is provided with the Coccinelle\footnote{\url{http://coccinelle.lip6.fr}} match and transformation engine. Semantic patch inference is well know in the realm of devide driver development where collateral evolutions can be summarized and specified in a patch that can change several code places simulateously~\cite{Andersen:2012:SPI:2351676.2351753,spdiff,bissyande2015implementing}. In our context, the main objective is to build upon the notion of semantic patch to find and represent common change patterns in several concrete patches from the fix examples yielded in previous step.
	
\begin{figure*}[t!]
		 \centering
        \begin{subfigure}[b]{0.5\textwidth}
                \includegraphics[width=\textwidth]{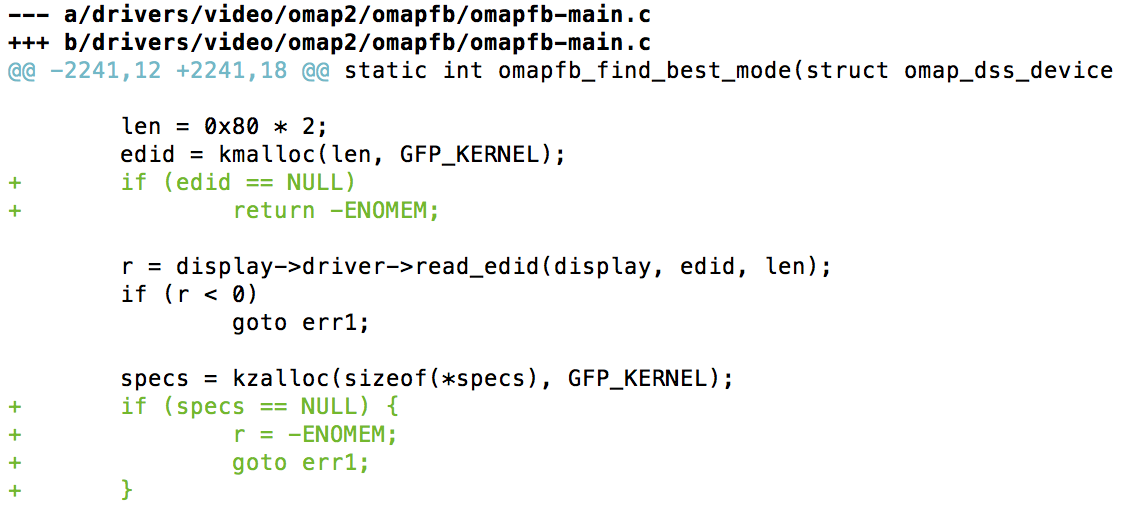}
                \caption{\scriptsize Linux commit {\tt fa0c5e71295fa4d62b900818d900c16980985e72}}
                \label{fig:gull}
        \end{subfigure}%
        ~ 
        \begin{subfigure}[b]{0.5\textwidth}
                \includegraphics[width=\textwidth]{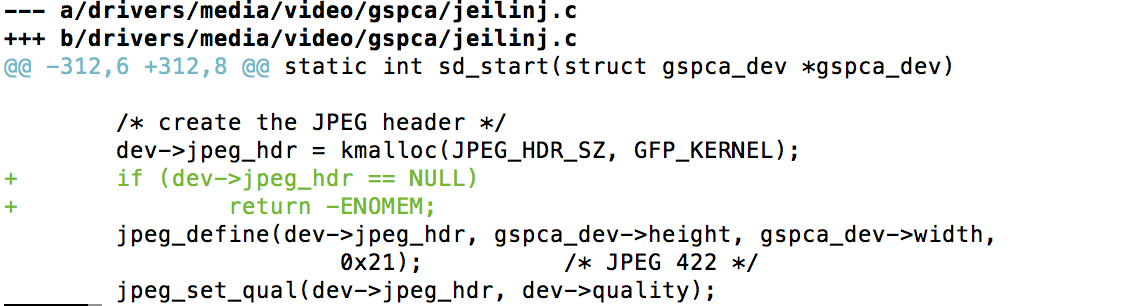}
                \caption{\scriptsize Linux commit {\tt d8370f7eff14ef646ba4a81eb9642800518a7324}}
                \label{fig:tiger}
        \end{subfigure}
        \caption{Concrete patches submitted to Linux kernel for preventing NULL pointer dereference faults}\label{fig:patches}
\end{figure*}

	In Figure~\ref{fig:patches} we illustrate concretes patches dealing with the common NULL pointer dereference fault in Linux kernel. In system code, memory allocation with {\em kmalloc} or {\em kzalloc} can fail and thus the pointer that was supposed to hold the memory is invalid. If in its code the developer attempts to access such invalid memory, the operating system crashes. Such a fault it typically avoided by checking the validity of the pointer and exiting the caller function with a memory error code ({\tt ENOMEM}).  
	
\begin{center}
	\begin{figure}[t!]
	 	\centering
 	\includegraphics[width=.8\linewidth]{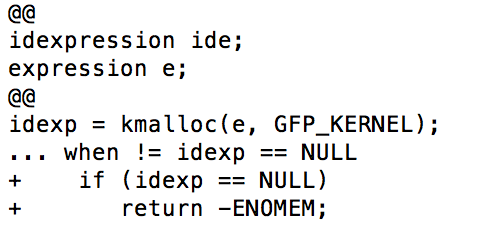}
 	\caption{Semantic patch summarizing the common fix actions of the concrete patches in Figure~\ref{fig:patches}}
 	\label{fig:cocci}
 \end{figure}
\end{center}

Figure~\ref{fig:cocci} illustrates a human-written example of semantic patch representing the common part of the concrete patches  of Figure~\ref{fig:patches}. In this code snipped we note that some meta-variables ({\em idexpression} and {\em expression}) are used to abstract over context-sensitive information. This semantic patch specifies that when a check is not performed on the pointer returned by kmalloc, one should add fix actions for checking this pointer and returning. Note that, although these actions should be performed also for kzalloc, the semantic patch does not take this into account as it focuses on the common change patterns.

\section{Preliminary Assessment}
Assessment of a recommender system is challenging since it should require a thorough evaluation via
a user study with developers. Instead for our preliminary assessment we propose
to empirically demonstrate that the proposed approach is simply sound. The main aim is to
show that with RECOMMEND, we can truly show fix tips that would have been
relevant to address bug reports. To that end, we separately investigate the different steps of the approach. Our experiments are performed on artefacts from Linux kernel development project.

First we consider the bug categorisation problem and evaluate with ten-fold cross validation whether RECOMMEND can predict with high accuracy and recall the category of a bug report and classify it with similar bugs. To that end, we manually collect a set of bug report discussing crash reports with kernel paging failures and NULL pointer dereference. We also consider other bugs that were obviously not related to NULL pointer dereference. We build our feature vectors with LDA topics to which each bug report belongs to then we applied 10-Fold cross validation on the classifiers built with SVM algorithm. We obtained a precision of 83\% with a recall at 74\%.

 Second we consider the collection of bug links for past bug reports. With the simple heuristic of greping the term ``{\tt Bug \#}'' in commit logs, we were able to collect around 400 bug reports with their links. Variations of this heuristic and use of more advanced bug linking strategies that recover completely missing links can substantially improve this step.
 
 Finally, for the summarization of fix hints, we consider the set of labelled bug reports used to evaluate the bug categorization step. We consider the subset of those bug reports that are referred to explicitely in code commit logs. Leveraging the spdiff\footnote{\url{http://www.diku.dk/~jespera/doc.html}} tool for semantic patch inference we compute the recurrence of fix actions. We found that over 50\% of the patches shared the a fix action checking the return value on {\em kmalloc}. This result is inline with findings in other studies about the repetitiveness of code changes.

\section{Discussion and On-Going Work}

We are currently preparing an extensive evaluation of the prototype implementation of RECOMMEND to take into account other software development projects, and a user study to assess the relevance of the fix tips recommended.

In the current form of the implementation, a number of choices have been made which contribute to yield a less optimum recommendation system. First, for the categorization of bug reports we have considered only the short description of the bug. In the field of Natural Language Processing, Rus et al. have already demonstrate similarity metrics based on Latent Dirichlet Allocation is also promising for short texts~\cite{Rus:2013:SMB:2468221.2468264}. Topic modeling however would be more reliable if we consider all bug report description and the associated comments. Unfortunately, our initial experiments have yielded bad results due to the nature of bug reports where code snippets and specific-environment execution outputs are included. In previous work, we have highlighted the same challenges which reduce the performance of bug linking approaches~\cite{Bissyande:2013:EEB:2495256.2495765}.

Second, for the inference of fix examples, we considered a basic linking strategies relying on the explicit ones. Yet, the state-of-the art literature contains various advanced approaches for bug linking. Unfortunately, the different strategies proposed only work for specific kinds of projects~\cite{Bissyande:2013:EEB:2495256.2495765}. Thus, there is no generic approach to bug linking to use in the RECOMMEND approach. To improve this step, we could devise a recommendation system for automatically selecting a bug linking strategy depending on the context. 

Third, our approach eludes the question of bug localization to help the developer fix the bug even more quickly. However, our approach builds on assumptions based on the findings of our previous studies~\cite{Bissyande2012} where we discuss for example the development of device drivers and how many faults occur when interacting with the API. Given this information and provided with a generic fix hints (which may contain a concrete term, e.g. the name of an API such as {\em kmalloc}) is already a big step forward.

Fourth, and finally, our current inference of semantic patch inference builds on top spdiff whose scalability is challenged. We are currently investigated improvements in the tool, in particular in a faster algorithm for pruning dissimilar parts of concrete patches to speed up the construction of the semantic patch for the common change patterns.

\section{Conclusion}\label{sec:conclusion}
In this paper we propose a motivated description of the RECOMMEND approach to recommending bug fixes. The main originality of our work is that: (1) it focuses on bug reports, which are actually the main means in industry for identifying bugs by relying on users to test the software in production; (2) it yields fix hints in a generic form so that developers can immediately understand what part of the fix is important (e.g., API concerned) and which part is actually context-dependant (e.g., variable names).

Preliminary assessment has revealed promising results. In future work we will address different challenges to improve the tool to better exploit the Natural language text in bug reports and the associated comments, as well to take advantage of existing bug linking approaches.

\let\oldthebibliography=\thebibliography
\let\endoldthebibliography=\endthebibliography
\renewenvironment{thebibliography}[1]{%
  \vspace{0pt}
  \begin{oldthebibliography}{#1}%
    \fontsize{7.0}{7.5}\selectfont
    \setlength{\parskip}{0ex}%
    \setlength{\itemsep}{0.5ex}%
}%
{%
  \end{oldthebibliography}%
}
{
\balance
\bibliographystyle{IEEEtran}

}

\end{document}